\documentclass[a4paper,12pt,fleqn]{article}

\usepackage[tbtags]{amsmath}
\usepackage{amssymb}
\usepackage{mathtools}

\newcommand{\ud}{\mathrm{d}}

\title{\bf Superunitary operator and BRST transformations for non-Abelian two-form}

\author{{\normalsize \textbf{Dheeraj Shukla, Kuldeep Kumar}}\\[0.0ex]
{\small \textit{Department of Physics, Panjab University, Chandigarh 160014, India}}\\[-0.4ex]
{\small \texttt{dheerajkumarshukla@gmail.com}, \texttt{kuldeepk@pu.ac.in}}
}

\date{}

\begin{document}

\maketitle

\begin{abstract}
Using superspace unitary operator formalism, we derive various \mbox{(anti-)BRST} symmetry transformations explicitly for the non-Abelian 2-form gauge theories. We introduce a new Lagrangian with a coupling of matter fields not only with 1-from background field but also with a 2-form field. Moreover, the two gauge fields couple mutually as well. A new covariant derivative involving the 2-form gauge field is introduced. We also put forth a conjecture to generalise this idea to any $p$-form gauge theory.
\end{abstract}


\section{Introduction}

The $p$-form objects are at the core of the higher form gauge theories. The gauge concept relies on 1-form in case of point particles. Just as 1-form (4-vector potential) couples to charged point particle, $p$-forms couple to higher-dimensional objects, such as strings, membranes, etc. A generalisation of the 1-form gauge theories to $p$-form gauge theories has been a long discussed problem in theoretical physics \cite{Teitelboim1986, HT1986, SOS1979, FT1981, CFT1995, CFT1996, Smolin2000, GV2001}. Since a 2-form couples to surface, it is natural to think of it as a gauge field for (open or closed) strings. Although this is a consistent picture for Abelian 2-form, it is problematic for its non-Abelian counterpart owing to the difficulties in defining surface-ordered exponentials, which appear when a 2-form is coupled to the world surface of a string \cite{Teitelboim1986}. Nevertheless, non-Abelian 2-forms have appeared in the context of nonlinear sigma model \cite{SOS1979, FT1981}, in the loop space formulations of Yang-Mills theory \cite{CFT1995, CFT1996} and gravity as a gauge theory \cite{Smolin2000, GV2001}.

Within the framework of BRST formalism,\footnote{BRST is the abbreviation of the names of its founders, viz.\ Becchi, Rouet, Stora and Tyutin.} one approach to $p$-form gauge theories is the superfield formalism \cite{BT1981, BPT1981, BPT1982, DJ1981, DJT1982}. A superfield is a function on superspace, which is a Minkowski spacetime augmented with additional Grassmann coordinates $\theta$ and $\bar\theta$. In the superfield formalism, (anti-)BRST transformations for the (non-)Abelian 1-form gauge and corresponding ghost and anti-ghost fields can be derived exploiting the so-called horizontality condition (HC). This condition is basically equating the supercurvature 2-form defined on $(D+2)$-dimensional superspace to the ordinary curvature 2-form defined on the $D$-dimensional Minkowski spacetime. To include interacting systems where the gauge field couples to matter fields, this formalism has been consistently generalised to obtain the (anti-)BRST transformations for the matter fields as well, which is called the augmented superfield formalism \cite{Malik2006, Malik2006b, Malik2006c, Malik2009}, where, in addition to the horizontality condition, some gauge-invariant restrictions (GIRs) are also exploited. The mapping of ordinary fields on the Minkowski spacetime to the superfields on the superspace can also be carried out via a superspace unitary operator \cite{BT1981, BPT1981, BPT1982, SBM2015, BSM2016}, or superunitary operator for short. The superunitary operator upgrades the fields and gauge connections to their superspace-counterparts in the same fashion as the unitary gauge operator maps the fields and gauge connections to their gauge-transformed counterparts. This superunitary operator is determined from the horizontality condition and gauge-invariant restrictions.

Our goal in this paper is to deduce the (anti-)BRST transformation for the Kalb-Raymond $\mathcal{B}$-field 2-form, following the superunitary operator approach. For that we consider the interacting theory where the matter fields interact with the 1-form as well as the 2-form gauge field. The two gauge fields, too, interact with each-other through the well known ${\cal B} \wedge {\cal F}$ interaction term. Our focus would be to find out the (anti-)BRST symmetry transformations for the various fields and to obtain the corresponding covariant derivatives for both the gauge fields, i.e.\ for 1-form and 2-form gauge connections.  

We start with a brief review, in Sec.~\ref{sec:one-form}, of 1-form gauge theories, (anti-)BRST transformations and the idea of superfields and superunitary operator. Exploiting the horizontality condition and the gauge-invariant restrictions, the superunitary operator is obtained which upgrades the fields to their superspace-counterparts. The expansion of the superfields in terms of the ordinary fields yields the (anti-)BRST transformations of that field. We have incorporated the matter fields also and both the Abelian and non-Abelian cases are discussed. In Sec.~\ref{sec:two-form}, we extend the superunitary operator formalism to 2-form non-Abelian gauge theories and obtain the (anti-)BRST transformations for the same. We also propose a conjecture to deal with the transformation of the $p$-form gauge field associated with the local scalar gauge symmetry of the theory. Our concluding remarks are left for Sec.~\ref{sec:conclu}.


\section{\label{sec:one-form}BRST transformations and superunitary operator formalism for 1-form gauge theories}

In this section we present a brief review of the superspace unitary operator formalism, which also helps to set up notations. We start with the usual 1-form electrodynamics and obtain the (anti-)BRST transformations for matter fields, gauge connection and the corresponding ghost and anti-ghost fields. The non-Abelian case is also discussed.

\subsection{Abelian 1-form gauge theories}

The world line for a point particle is a 1-dimensional object in the background sapcetime. If $\tau$ is the parameter of the world line then the tangent to the particle trajectory is given as $u^\mu = {\ud{x^\mu}(\tau)/ {\ud\tau}}$. The action for a charged particle, of charge $q$ and mass $m$, in the presence of interacting background is given by
\begin{equation}
\begin{aligned}
S &= -\,m\int \ud\tau + q\oint {\cal A} -\,\frac{1}{2}\,\int \ud^4 x \sqrt{-g}\, ({\cal F} \wedge *{\cal F})\\
&= -\,m\int \ud\tau + q\oint {\cal A}_\mu\,{u^\mu}\, \ud\tau -\,\frac{1}{4}\,\int \ud^4 x \sqrt{-g}\, {\cal F}_{\mu\nu}\,{\cal F}^{\mu\nu},
\end{aligned}
\end{equation}
where ${\cal A} = {\cal A}_\mu (x)\, \ud x^\mu$ is the 1-form gauge field interacting with the charged particle and ${\cal F} = \ud \mathcal{A}= (1/2) {\cal F}_{\mu\nu}\, \ud x^\mu \wedge \ud x^\nu$ is the corresponding 2-form curvature field: ${\cal F}_{\mu\nu} = \partial_\mu {\cal A}_\nu - \partial_\nu {\cal A}_\mu $. If we go through the properties of the gauge field, which is a connection 1-form appearing due to an interaction of the charged particle with the background spacetime, we find that this is nothing but the well-studied electromagnetic field. The Lagrangian density for this background field,
\begin{equation}
{\cal L}_0 = -\,\frac{1}{4}{\cal F}_{\mu\nu}\,{\cal F}^{\mu\nu},
\end{equation}
remains invariant under the gauge transformation
\begin{equation}
\mathcal{A}_\mu \rightarrow  \mathcal{A}'_\mu = \mathcal{A}_\mu + \partial_\mu \chi,
\end{equation}
where $\chi (x)$ is some arbitrary scalar field. This Lagrangian density for the electromagnetic field is not easy to quantise because of the spurious degrees of freedom present in it. One of the best remedies to quantise is to use the BRST formalism.\footnote{We shall not go into the details of quantisation as it is not our concern here.}

The (anti-)BRST symmetry invariant off-shell 
 Lagrangian density can be written as
\begin{equation}\label{l1}
{\cal L}_1 = -\,\frac{1}{4}\,{\cal F}_{\mu\nu}{\cal F}^{\mu\nu} + \frac{1}{2}{B^2} + B(\partial_\mu {\cal A}^\mu) - i\partial_\mu \bar C \,\partial^\mu C,
\end{equation}
where $B(x)$ is the Nakanishi-Lautrup auxiliary scalar field, introduced to linearise the Feynman gauge fixing term: $-\,(1/2)(\partial_\mu {\cal A}^\mu)^2$. This also helps to obtain an off-shell nilpotency for the Fadeev-Popov (anti-)ghost fields $(\bar C)C$ which appear in 1-loop Feynman diagram for the electromagnetic theory. These (anti-)ghost fields are virtual scalar fields but they satisfy the Grassmann-odd properties:
\begin{equation}
C^2 = 0, \quad \bar C^2 = 0, \quad \bar C C + C \bar C = 0.
\end{equation}
The Lagrangian density \eqref{l1} is invariant under the following (anti-)BRST transformations:
\begin{equation}
\begin{aligned}
&s_b\, {\cal A}_\mu = \partial_{\mu} C, \quad s_b C = 0, \quad s_b \bar C = iB, \quad s_b  B = 0, \\
&s_{ab}\, {\cal A}_\mu = \partial_{\mu} \bar C, \quad s_{ab} \bar C = 0, \quad s_{ab}  C = -i B, \quad s_{ab} B = 0.
\end{aligned}
\end{equation}
These transformations are anti-commuting, $s_b s_{ab} + s_{ab} s_b = 0$, and off-shell nilpotent, $s_b^2 = s_{ab}^2 = 0$.

Next, we consider the 1-form Abelian electrodynamics:
\begin{equation}\label{l2}
\mathcal{L}_2 = -\frac{1}{4} \mathcal{F}_{\mu\nu} \mathcal{F}^{\mu\nu} + \bar{\psi} (i\gamma^\mu D_\mu - m)\psi,
\end{equation}
where $D_\mu$ is the covariant derivative: $D_\mu \psi = \partial_\mu \psi + i \mathcal{A}_\mu \psi$. This Lagrangian density is invariant under the $U(1)$ gauge symmetry transformations ($q=1$)
\begin{equation}\label{gauge-trans1}
\begin{aligned}
&\psi(x)\to \psi'(x) = U(x)\psi (x),\quad \bar\psi(x) \to \bar\psi'(x) = \bar\psi(x)U^\dagger (x),\\
&\mathcal{A}_\mu (x)\to \mathcal{A}'_\mu(x) = U(x) \mathcal{A}_\mu(x)\,U^\dagger (x) + i (\partial_\mu U) U^\dagger,
\end{aligned}
\end{equation}
where $U(x) = \exp (- i \chi(x))$. Obviously, the operator $U$ is unitary: $UU^\dagger = 1 = U^\dagger U$. The infinitesimal version of these transformations is
\begin{equation}
\begin{aligned}
&\psi(x)\to \psi'(x) = \psi (x) - i \chi \psi,\quad \bar\psi(x) \to \bar\psi'(x) = \bar\psi(x) + i \bar\psi \chi,\\
&\mathcal{A}_\mu \rightarrow  \mathcal{A}'_\mu = \mathcal{A}_\mu + \partial_\mu \chi.
\end{aligned}
\end{equation}
The (anti-)BRST invariant version of the Lagrangian density \eqref{l2} is 
\begin{equation}\label{l3}
\mathcal{L}_3 =  -\,\frac{1}{4}\, {\cal F}_{\mu\nu}{\cal F}^{\mu\nu} + \bar\psi (i\gamma^\mu D_\mu - m)\,\psi + \frac{1}{2}{B^2} + B(\partial_\mu {\cal A}^\mu) - \, i\,\partial_\mu\bar C\partial^\mu C,
\end{equation}
which is invariant under the following (anti-)BRST symmetry transformations:
\begin{equation}\label{brst-trans}
\begin{aligned}
&s_b {\cal A}_\mu = \partial_\mu C, \quad s_b C = 0, \quad s_b \bar C = i B, \quad s_b B = 0,\\
&s_b \psi = -i C\psi, \quad s_b\bar\psi = -i \bar\psi C,\\
&s_{ab} {\cal A}_\mu = \partial_\mu \bar C, \quad s_{ab}\bar C = 0, \quad s_{ab} C = -i B, \quad s_{ab} B = 0,\\
&s_{ab} \psi = -i \bar C\psi, \quad s_{ab}\bar\psi = -i\bar\psi \bar C.
\end{aligned}
\end{equation}

\subsection{Superunitary operator formalism}

The symmetry transformations \eqref{brst-trans} can be obtained using a number of methods, viz.\ the usual constraint analysis method \cite{Sundermeyer1982}, the augmented superfield formalism, etc.\ \cite{Teitelboim1986, HT1986, Malik2006}. The superfield formalism relies on the extension of the ordinary spacetime to a superspace which has a space of Grassmann-odd coordinates, $\theta$ and $\bar\theta$ (satisfying $\theta^2 = {\bar\theta}^2 = 0$, $\theta \bar\theta + \bar\theta \theta =0$), attached to each and every point of ordinary spapcetime. Any field or operator in ordinary spacetime gets upgraded to its superspace-counterpart. A general expansion of the spinor fields is written as
\label{psi-map}
\begin{gather}\label{psi-map1}
\psi(x) \to \Psi(x, \theta, \bar\theta) = \psi (x) + \theta\,\bar p(x) + \bar\theta\, p(x) + \theta\bar\theta\, q(x), \\
\label{psi-map2}
\bar\psi(x) \to {\overline \Psi}(x, \theta, \bar\theta) = \bar\psi (x) + \theta\,\bar r(x) + \bar\theta\, r(x) + \theta\bar\theta\, t(x),
\end{gather}
where the auxiliary fields $p$, $\bar p$, $r$ and $\bar r$ have Grassmann-even character while the auxiliary fields $q$ and $t$ are of the Grassmann-odd character. The 1-form gauge field ${\cal A}$ is extended as
\begin{equation}\label{a-map}
\begin{split}
{\cal A}(x)\to {\cal \cal\tilde A}(x, \theta, \bar\theta) &=  \ud x^\mu\, E_\mu (x, \theta, \bar\theta) + \ud \theta\, \bar F(x, \theta, \bar\theta) + \ud \bar\theta\, F(x, \theta, \bar\theta) \\
&=  E(x, \theta, \bar\theta) + \ud\theta\, \bar F(x, \theta, \bar\theta) + \ud\bar\theta\, F(x, \theta, \bar\theta),
\end{split}
\end{equation}
where the superfields $E$, $F$ and $\bar F$ can be further decomposed as
\begin{equation}\label{effbar1}
\begin{aligned}
&E(x, \theta, \bar\theta) = {\cal A}(x) + \theta \,\bar R(x) + \bar\theta \,R(x) + \theta\bar\theta \,H(x), \\
&F(x, \theta, \bar\theta) =  C(x) + \theta\,\bar K(x) + \bar\theta \, K(x) + \theta\bar\theta\, S(x), \\
&\bar F(x, \theta, \bar\theta) = \bar C(x) + \theta\,\bar L(x) + \bar\theta\,L(x) + \theta\bar\theta\, T(x). 
\end{aligned}
\end{equation}
Obviously, the auxiliary fields $R$, $\bar R$, $S$ and $T$ have Grassmann-odd character while the fields $H$, $K$, $\bar K$, $L$ and $\bar L$ have Grassmann-even character. 

Now we shall adopt a more intuitive approach, the superunitary operator formalism \cite{BT1981, BPT1981, BPT1982, SBM2015, BSM2016}. The philosophy of this approach is that a superunitry operator upgrades the fields and gauge connections on ordinary spacetime to their counterparts on the superspace in the same fashion as the unitary operator upgrades the fields and gauge connection to their gauge-transformed counterparts on the ordinary spacetime. Thus, the gauge transformations \eqref{gauge-trans1} dictate the upgradation of $\psi$, $\bar\psi$ and $\mathcal{A}$ as
\begin{equation}\label{upgrade1}
\begin{aligned}
&\psi(x) \to \Psi(x, \theta, \bar\theta) = \tilde{U} (x, \theta, \bar\theta)\,\psi (x),\quad
\bar\psi(x) \to \overline{\Psi}(x, \theta, \bar\theta) = \bar\psi (x)\,\tilde{U}^\dagger (x, \theta, \bar\theta), \\
&{\cal A}(x)\to {\cal \tilde A}(x, \theta, \bar\theta) =  \tilde{U} (x, \theta, \bar\theta)\,{\cal A}(x)\,\tilde{U}^\dagger (x, \theta, \bar\theta) + \tilde\phi (x, \theta, \bar\theta),
\end{aligned}
\end{equation}
where $\tilde\phi (x, \theta, \bar\theta) = i\tilde \ud \tilde{U}(x, \theta, \bar\theta)\;\tilde{U}^\dagger (x, \theta, \bar\theta) $ and $\tilde \ud = \ud x^\mu \partial_\mu + \ud \theta\partial_\theta + \ud \bar\theta\partial_{\bar\theta}$ is the extension of the ordinary exterior derivative $\ud = \ud x^\mu \partial_\mu$. The superunitary operator $\tilde{U}$ will now be obtained using the horizontality condition and the gauge invariant restrictions.

First we exploit the honizontality condition\footnote{The horizontality condition relies on the fact that a physical field is independent of the Grassmann variables, i.e.\ $\theta$ and $\bar\theta$. Such a physical field, $\mathcal{F}$ for example, transforms as $\mathcal{F} \to \tilde{\mathcal{F}} = \tilde{U} \mathcal{F} \tilde{U}^\dagger$. For the Abelian case, $\tilde{\mathcal{F}} = \tilde{U} \mathcal{F} \tilde{U}^\dagger = \mathcal{F} \Rightarrow \tilde{\ud}\tilde{\mathcal{A}} = \ud \mathcal{A}$.}
\begin{equation}\label{hc}
\tilde \ud\tilde{\cal A} = \ud{\cal A}.
\end{equation} 
In view of the relations,
\begin{equation}
\begin{aligned}
&\ud\theta \wedge \ud\bar\theta = \ud\bar\theta \wedge \ud\theta, \quad \ud\theta \wedge \ud\theta \neq 0,\quad \ud\bar\theta \wedge \ud\bar\theta \neq 0, \\ 
&\lbrace \ud\theta, \partial_\theta \rbrace = 0, \quad \lbrace \ud\bar\theta, \partial_{\bar\theta} \rbrace = 0,\quad \lbrace \ud\theta, \partial_{\bar\theta} \rbrace = 0,\quad \lbrace \ud\bar\theta, \partial_\theta\rbrace = 0, \\
&\ud x^\mu \wedge \ud\theta = - \ud\theta\wedge \ud x^\mu, \quad \ud\bar\theta \wedge \ud x^\mu = - \ud x^\mu\wedge \ud\bar\theta,\quad \ud x^\mu \wedge \ud x^\nu = - \ud x^\nu\wedge \ud x^\mu,
\end{aligned}
\end{equation}
the honizontality condition \eqref{hc} reduces to 
\begin{equation}
\begin{split}
\ud \mathcal{A} &= \ud E + \ud\theta \wedge (\partial_\theta E - \ud\bar F) + \ud\bar\theta \wedge (\partial_{\bar\theta}E - \ud F) \\
&{}\quad - (\ud\theta \wedge \ud\theta)\,\partial_\theta \bar F - (\ud\bar\theta \wedge \ud\bar\theta)\,\partial_{\bar\theta} F - (\ud\theta \wedge \ud\bar\theta)\,(\partial_\theta F + \partial_{\bar\theta} \bar F).
\end{split}
\end{equation}
Comparing the coefficients of various differentials on both the sides, we get
\begin{equation}
\begin{aligned}
&\ud E = \ud{\cal A}, \quad \partial_\theta E = \ud\bar F,\quad \partial_{\bar\theta} E = \ud F, \\
&\partial_\theta \bar F = 0,\quad \partial_{\bar\theta} F = 0, \quad \partial_\theta F + \partial_{\bar\theta} \bar F = 0,
\end{aligned}
\end{equation}
which, using \eqref{effbar1} yields $\bar L = 0$, $T=0$, $K=0$, $S=0$, $\bar K = -L$, $R=\ud C$, $\bar R = \ud \bar C$, $H=\ud L$. Thus equation \eqref{effbar1} reduces to
\begin{equation}\label{effbar2}
\begin{aligned}
&E (x, \theta, \bar\theta) = {\cal A} (x) + \theta\, \ud\bar C(x) + \bar\theta\, \ud C (x) + \theta\bar\theta\, \ud L (x), \\
&F (x, \theta, \bar\theta) = C(x) - \theta L(x), \\
&\bar F (x, \theta, \bar\theta) = \bar C (x) + \bar\theta L (x).
\end{aligned}
\end{equation}
From the relations \eqref{effbar1}, or \eqref{effbar2}, we notice the following correspondence between the fields on ordinary spacetime and their superspace-counterparts: ${\cal A}(x) \to E(x, \theta, \bar\theta)$, $C(x) \to F(x, \theta, \bar\theta)$, $\bar C(x) \to \bar F(x, \theta, \bar\theta)$. Since $BA + i \bar C \ud C$ is a gauge-invariant quantity, as can be easily verified using \eqref{brst-trans}, we now exploit the following gauge invariant restriction:\footnote{A gauge-invariant restriction is the expression which retains the same form while upgrading from ordinary spacetime to superspace.}
\begin{equation}
B(x) \,E(x, \theta, \bar\theta)  + i\,\bar F (x, \theta, \bar\theta)\,  \ud F (x, \theta, \bar\theta) =  B (x)\,{\cal A}(x) + i\,\bar C (x)\, \ud C(x),
\end{equation}
which fixes $L = i B$. Therefore, the relations \eqref{effbar2} reduce to
\begin{equation}\label{effbar3}
\begin{aligned}
&E (x, \theta, \bar\theta) = {\cal A} (x) + \theta\, \ud\bar C(x) + \bar\theta\, \ud C (x) + i\theta\bar\theta\, \ud B (x), \\
&F (x, \theta, \bar\theta) = C(x) - i \theta B(x), \\
&\bar F (x, \theta, \bar\theta) = \bar C (x) + i \bar\theta B (x),
\end{aligned}
\end{equation}
which are the final expressions for the superfields $E$, $F$ and $\bar F$, in terms of the basic known fields.

As mentioned earlier in \eqref{upgrade1}, the spinor field upgrades to its superspace-counterpart as $\psi(x) \to \Psi(x, \theta, \bar\theta) = \tilde{U} (x, \theta, \bar\theta)\,\psi (x)$. The covariant derivative $D \psi = (\ud + i\mathcal{A})\psi$ should also, therefore, transforms in the same fashion,
\begin{equation}\label{dpsi}
D\psi(x) \to \tilde{D}\Psi(x, \theta, \bar\theta) = \tilde{U} (x, \theta, \bar\theta)\,D\psi (x),
\end{equation}
where 
\begin{equation}\label{dtil}
\tilde D =  \tilde \ud + i \tilde {\cal A} = \ud + \ud \theta\partial_\theta + \ud \bar\theta\partial_{\bar\theta} + i \tilde {\cal A}.
\end{equation}
Using \eqref{dtil} and \eqref{psi-map1}, we now evaluate $\tilde{D} \Psi$, which in view of \eqref{a-map} and \eqref{effbar3} gives
\begin{equation}\label{dpsi0}
\begin{split}
\tilde{D} \Psi(x, \theta, \bar\theta) &= (\ud + i\mathcal{A} + i\theta\ud \bar C + i \bar\theta \ud C - \theta\bar\theta \ud B)\psi \\
&{} \quad + (\ud + i\mathcal{A})\theta\bar p + (\ud + i\mathcal{A})\bar\theta p + (\ud + i\mathcal{A})\theta\bar\theta q \\
&{} \quad + \ud \theta\, [\bar p + i\bar C \psi + \theta (-i \bar C \bar p) + \bar\theta (q - i \bar C p -B \psi) + \theta\bar\theta (i\bar C q + B\bar p)] \\
&{} \quad + \ud \bar\theta\, [p + i C \psi + \bar\theta (iCp) + \theta (-q -i C \bar p + B\psi) + \theta\bar\theta (i Cq + Bp)].
\end{split}
\end{equation}
In view of \eqref{dpsi}, we now equate the right-hand side of \eqref{dpsi0} to $\tilde{U}D\psi = \tilde{U}(\ud + i\mathcal{A})\psi$, which gives
\begin{gather}
\label{dpsi1}
\begin{split}
\tilde{U}  (\ud + i\mathcal{A}) \psi &= (\ud + i\mathcal{A} + i\theta\ud \bar C + i \bar\theta \ud C - \theta\bar\theta \ud B)\psi \\
&{} \quad + (\ud + i\mathcal{A})\theta\bar p + (\ud + i\mathcal{A})\bar\theta p + (\ud + i\mathcal{A})\theta\bar\theta q,
\end{split} \\
\label{dpsi2}
(\bar p + i\bar C \psi) + \theta (-i \bar C \bar p) + \bar\theta (q - i \bar C p -B \psi) + \theta\bar\theta (i\bar C q + B\bar p) =0, \\
\label{dpsi3}
(p + i C \psi) + \bar\theta (iCp) + \theta (-q -i C \bar p + B\psi) + \theta\bar\theta (i Cq + Bp) =0.
\end{gather}
Setting the terms inside the parentheses on the left-hand sides of \eqref{dpsi2} and \eqref{dpsi3} individually to zero, fixes $p = -i C\psi$, $\bar p = -i \bar C \psi$, $q= (B-C \bar C)\psi$, which reduces \eqref{psi-map1} to
\begin{equation}\label{psi-map11}
\Psi(x, \theta, \bar\theta) = \left[1 - i\theta \bar C - i\bar\theta C + \theta\bar\theta (B + \bar C C)\right] \psi(x),
\end{equation}
and \eqref{dpsi1} to
\begin{equation}
\tilde{U} D \psi = \left[1 - i\theta \bar C - i\bar\theta C + \theta\bar\theta (B + \bar C C)\right] D \psi(x),
\end{equation}
We can therefore identify the superunitary operator $\tilde U (x, \theta, \bar\theta)$ as
\begin{equation}\label{u}
\begin{split}
\tilde U (x, \theta, \bar\theta) &= 1 - i\theta \bar C - i\bar\theta C + \theta\bar\theta (B + \bar C C) \\
&= \exp [- i(\theta \bar C + \bar\theta C + i\theta\bar\theta B)].
\end{split}
\end{equation}
Using \eqref{psi-map2} and \eqref{psi-map11} and exploiting the gauge-invariant restriction
\begin{equation}
\overline{\Psi} (x, \theta, \bar\theta)\Psi  (x, \theta, \bar\theta) = \bar\psi(x)\psi(x),
\end{equation}
we get $r = -i \bar \psi C$, $\bar r = -i \bar\psi \bar C$, $t= -\bar\psi(B+C \bar C)$, which reduces \eqref{psi-map2} to
\begin{equation}\label{psi-map21}
\overline \Psi(x, \theta, \bar\theta) = \bar \psi \left[1 + i\theta \bar C + i\bar\theta C - \theta\bar\theta (B - \bar C C)\right].
\end{equation}
Comparing the above equation with $\overline{\Psi}(x, \theta, \bar\theta) = \bar\psi (x)\,\tilde{U}^\dagger (x, \theta, \bar\theta)$ then gives
\begin{equation}\label{udagger}
\begin{split}
{\tilde U}^\dagger (x, \theta, \bar\theta) &= 1 + i\theta \bar C + i\bar\theta C - \theta\bar\theta (B - \bar C C) \\
&= \exp [i(\theta \bar C + \bar\theta C + i\theta\bar\theta B)].
\end{split}
\end{equation}
As expected, the superunitary operator $\tilde U$ satisfies the unitarity condition, $ \tilde U \,\tilde U^\dagger = 1 = \tilde U^\dagger\,\tilde U $. Finally, comparing equations \eqref{effbar3}, \eqref{psi-map11} and \eqref{psi-map21} with the generic expansion of a superfield $G (x,\theta,\bar\theta)$ in terms of its ordinary counterpart $G(x)$,
\begin{equation}\label{generic}
G (x, \theta, \bar\theta)  = G (x) +  \theta\,(s_{ab}G)(x) + \bar\theta\,(s_b G)(x) + \theta\bar\theta\,(s_b\,s_{ab}G)(x),
\end{equation}
and keeping in mind the correspondence ${\cal A}(x) \to E(x, \theta, \bar\theta)$, $C(x) \to F(x, \theta, \bar\theta)$, $\bar C(x) \to \bar F(x, \theta, \bar\theta)$, $\psi(x) \to \Psi(x, \theta, \bar\theta)$, $\bar\psi(x) \to \overline\Psi(x, \theta, \bar\theta)$, the following (anti-)BRST transformations follow:
\begin{equation}\label{brst-trans2}
\begin{aligned}
&s_b {\cal A} = \ud C, \quad s_b C = 0, \quad s_b \bar C = i B, \quad
s_b \psi = -i C\psi, \quad s_b\bar\psi = -i \bar\psi C,\\
&s_{ab} {\cal A} = \ud \bar C, \quad s_{ab}\bar C = 0, \quad s_{ab} C = -i B, \quad
s_{ab} \psi = -i \bar C\psi, \quad s_{ab}\bar\psi = -i\bar\psi \bar C.
\end{aligned}
\end{equation}
Also, since $s_b^2 = 0$ we see that $s_b(s_b \bar C) = s_b (iB) = 0$, which implies $s_b B = 0$. Similarly it follows that $s_{ab} B = 0$. This completes the obtention of (anti-)BRST transformations \eqref{brst-trans} using the superunitary operator formalism.

\subsection{Non-Abelian 1-form gauge theories}

The complete gauge invariant Lagrangian density for the 1-form non-Abelian gauge field can be written as%
\footnote{The $\mathrm{SU}(N)$ generators $T^a$ (with $a = 1,2,..., N^2 - 1$) satisfy the commutation relation $[T^a, T^b] =  i\,f^{abc}\, T^c$, where summation over the repeated index is implied and the structure constants $f^{abc}$ are chosen to be totally antisymmetric in indices. The dot and cross products between two  vectors $P=P^a T^a$ and $Q=Q^a T^a$ are defined as $P \cdot Q = P^a Q^a,\; P \times Q = (f^{abc} P^b Q^c) T^a = -i[P,Q]$.}
\begin{equation}\label{l4}
\begin{split}
{\cal L}_4 &= \bar\psi (i\gamma^\mu D_\mu - m)\,\psi -\,\frac{1}{4}\, {\cal F}_{\mu\nu}\cdot { \cal F}^{\mu\nu} + \frac{1}{2}(B\cdot B + \bar B \cdot \bar B)\\
&\quad{} + B\cdot(\partial_\mu {\cal A}^\mu) - \, i\,\partial_\mu \bar C\cdot D^\mu C,
\end{split}
\end{equation}
where $\mathcal{F}_{\mu\nu} = \partial_\mu \mathcal{A}_\nu - \partial_\nu \mathcal{A}_\mu + i [\mathcal{A}_\mu, \mathcal{A}_\nu] = \partial_\mu \mathcal{A}_\nu - \partial_\nu \mathcal{A}_\mu - \mathcal{A}_\mu \times \mathcal{A}_\nu$ and the covariant derivatives are defined as $D_\mu \psi = \partial_\mu \psi + i \mathcal{A}_\mu \psi$, $D_\mu C = \partial_\mu C + i\,[{\cal A}_\mu, C] = \partial_\mu C - {\cal A}_\mu \times C$. A coupled but equivalent Lagrangian density can be obtained from the above Lagrangian density by using the celebrated Curci-Ferrari condition, $B + \bar B + i(\bar C \times C) = 0$, as%
\footnote{Owing to the Grassmann nature or $C$ and $\bar C$, the following relations hold: $\bar C \times C = C \times \bar C = -i \{C,\bar C\}$, $C \times C = -2i CC$, $C \cdot \bar C = -\bar C \cdot C$. It is also perhaps worthwhile to mention here that in the Abelian case $\bar C \times C = 0$, in which case the Curci-Ferrari condition reduces to $B = -\bar B$. Then the Lagrangian densities \eqref{l4} and \eqref{l5} reduce to that of the Abelian case, Eq.~\eqref{l3}.}
\begin{equation}\label{l5}
\begin{split}
{\cal L}_5 &= \bar\psi (i\gamma^\mu D_\mu - m)\,\psi -\,\frac{1}{4}\, {\cal F}_{\mu\nu}\cdot { \cal F}^{\mu\nu} + \frac{1}{2}(B\cdot B + \bar B \cdot \bar B)\\
&\quad {} - \bar B\cdot(\partial_\mu {\cal A}^\mu) - \, i\,D_\mu \bar C\cdot \partial^\mu C.
\end{split}
\end{equation} 
It can be shown that the action corresponding to the above Lagrangian densities remains invariant under the following (anti-)BRST transformations:
\begin{equation}\label{brst-trans-na}
\begin{aligned}
&s_b {\cal A}_\mu = D_\mu C, \quad s_b C = -i CC, \quad s_b \bar C = i B, \quad s_b B = 0, \quad s_b \bar B = i [\bar B, C], \\
&s_b \psi = -i C\psi, \quad s_b\bar\psi = -i \bar\psi C,\\
&s_{ab} {\cal A}_\mu = D_\mu \bar C, \quad s_{ab}\bar C = -i \bar C \bar C, \quad s_{ab} C = i \bar B, \quad s_{ab} \bar B = 0,\quad s_{ab}\, B = i [B, \bar C]\\
&s_{ab} \psi = -i \bar C\psi, \quad s_{ab}\bar\psi = -i\bar\psi \bar C.
\end{aligned}
\end{equation}

Now we shall demonstrate how these (anti-)BRST transformations can be obtained using the superunitary operator formalism. Instead of redoing all the steps of the Abelian case now for the non-Abelian case, it is more convenient to focus on the form of the superunitary operator first. We generalise Eqs.~\eqref{u} and \eqref{udagger} to
\begin{equation}
\begin{aligned}
&\tilde U (x, \theta, \bar\theta) = 1 - i\,(\theta\bar C + \bar\theta C) + \theta\bar\theta\,(f(\bar C C) + B)), \\
&\tilde U^\dagger(x, \theta, \bar\theta) = 1 + i\,(\theta\bar C + \bar\theta C) - \theta\bar\theta\,(f^\dagger (\bar C C) + B)),
\end{aligned}
\end{equation}
where $f$ and $f^\dagger$ are some combinations of $C\bar C$ and $\bar C C$. Now demanding the unitarity condition, we obtain $f = \bar C C$ and $f^\dagger = C \bar C$. With this, the final form of superunitary operators turns out to be
\begin{equation}\label{u-na}
\begin{aligned}
&\tilde U (x, \theta, \bar\theta) = 1 - i\,(\theta\bar C + \bar\theta C) + \theta\bar\theta\,(\bar C C + B),\\
&\tilde U^\dagger(x, \theta, \bar\theta) = 1 + i\,(\theta\bar C + \bar\theta C) - \theta\bar\theta\,(C\bar C + B).
\end{aligned} 
\end{equation}
As done earlier for the Abelian case in Eq.~\eqref{upgrade1}, the fields $\psi$, $\bar\psi$ and $\mathcal{A}$ are now upgraded to their superspace counterparts with help of the superunitary operator as
\begin{equation}\label{upgrade1-na}
\begin{aligned}
&\Psi(x, \theta, \bar\theta) = \tilde{U} (x, \theta, \bar\theta)\,\psi (x),\quad
\overline{\Psi}(x, \theta, \bar\theta) = \bar\psi (x)\,\tilde{U}^\dagger (x, \theta, \bar\theta), \\
&{\cal \tilde A}(x, \theta, \bar\theta) =  \tilde{U} (x, \theta, \bar\theta)\,{\cal A}(x)\,\tilde{U}^\dagger (x, \theta, \bar\theta) + \tilde\phi (x, \theta, \bar\theta).
\end{aligned}
\end{equation}
Thus using the form \eqref{u-na} in \eqref{upgrade1-na} we find
\begin{gather}\label{psi-map11-na}
\Psi(x, \theta, \bar\theta) = \left[1 - i\theta \bar C - i\bar\theta C + \theta\bar\theta (B - C\bar C)\right] \psi(x), \\
\label{psi-map21-na}
\overline \Psi(x, \theta, \bar\theta) = \bar \psi (x) \left[1 + i\theta \bar C + i\bar\theta C - \theta\bar\theta (B + C\bar C)\right], \\
\label{a-map-na}
\begin{split}
\tilde{\mathcal{A}}(x, \theta, \bar\theta) &= {\cal A} - \theta\,({\cal A}\times \bar C) - \bar\theta\,({\cal A}\times C)
-i \theta\bar\theta\,\left( {\cal A} \times B + i\, ({\cal A}\times C)\times \bar C \right) \\
&{}\quad + \left(\theta\, \ud\bar C + \bar\theta \, \ud C + i\theta\bar\theta\,(\ud B + \lbrace \ud C, \bar C\rbrace)\right) \\
&{}\quad + \ud\theta\,\left(\bar C - i\theta\,\bar C\bar C + i\bar\theta\,B + \theta\bar\theta\,[B, \bar C] \right) \\
&{}\quad + \ud\bar\theta \,\left( C - i\theta\,(B + \lbrace C, \bar C\rbrace) - i\bar\theta\, CC + 
\theta\bar\theta\,([B,C] + [CC, \bar C])\right).
\end{split}
\end{gather}
In view of \eqref{a-map}, we compare Eq.~\eqref{a-map-na} with ${\cal \cal\tilde A}(x, \theta, \bar\theta) =  E(x, \theta, \bar\theta) + \ud\theta\, \bar F(x, \theta, \bar\theta) + \ud\bar\theta\, F(x, \theta, \bar\theta)$, which yields
\begin{gather}
\label{e-na}
E =  {\cal A} + \theta\,(D\bar C) + \bar\theta \,(DC) + i\theta\bar\theta\,\left(DB + i\, \bar C\times DC\right),\\
\label{f-na}
F  =  C + i\theta\,\bar B + \frac{1}{2}\bar\theta\, \left( C\times C\right) - i\theta\bar\theta\,(\bar B\times C), \\
\label{fbar-na}
\bar F =  \bar C + \frac{1}{2}\theta\,\left( \bar C \times \bar C \right) + i\bar\theta\,B 
+ i\, \theta\bar\theta\,(B \times \bar C),
\end{gather}
where we have also made use of the Curci-Ferrari condition, $B + \bar B + i\, (\bar C \times C) = 0$. Now the comparison of Eqs.~\eqref{psi-map11-na}, \eqref{psi-map21-na}, \eqref{e-na}--\eqref{fbar-na} with \eqref{generic} yields the (anti-)BRST transformations \eqref{brst-trans-na}.


\section{\label{sec:two-form}BRST transformations for non-Abelian 2-form gauge field}

A string is a 1-dimensional object which when moves in the background spacetime, traces out a 2-dimensional surface in spacetime. The action for an interacting string, in the background of a Kalb-Raymond 2-form gauge field ${\cal B} = \tfrac{1}{2} (\ud x^\mu \wedge \ud x^\nu)\; {\cal B}_{\mu\nu}$, can be written as \cite{Zwiebach2009}
\begin{equation}
\begin{split}
S &=  S_f - \,\oint {\cal B} - \frac{1}{2} \,\int \sqrt{-g}\, \ud^4 x\; {\cal H}\wedge *{\cal H} \\
&=  S_f - \frac{1}{2!}\,\oint \ud\tau \ud\sigma \;{\cal B}_{\mu\nu} (X (\tau,\sigma))\,\frac{\ud X^{[\mu}}{\ud\tau}\, \frac{\ud X^{\nu]}}{\ud\sigma} \\
&\quad {} - \frac{1}{2\,(3!)}\,\int \sqrt{-g}\, \ud^4 x\;{\cal H}_{\mu\nu\lambda}\,{\cal H}^{\mu\nu\lambda},
\end{split}
\end{equation}
where $S_f$ is the free string action and ${\cal H} = \ud \mathcal{B}= \frac{1}{3!}(\ud x^\mu \wedge \ud x^\nu \wedge \ud x^\lambda)\,{\cal H}_{\mu\nu\lambda}$, with ${\cal H}_{\mu\nu\lambda} = \partial_\mu{\cal B}_{\nu\lambda} + \partial_\nu{\cal B}_{\lambda\mu} + \partial_\lambda{\cal B}_{\mu\nu}$, is the curvature 3-form corresponding to the ${\cal B}$-field, while $M^{[\mu}N^{\nu]}$ is a shorthand notation for $M^\mu N^\nu - N^\nu M^\mu$. In light of this, the Kalb-Raymond Lagrangian density, for the non-Abelian case, can be written as
\begin{equation}
{\cal L}_6 = \frac{1}{12}\, {\cal H}_{\mu\nu\lambda} \cdot{\cal H}^{\mu\nu\lambda}
\end{equation}
This is one way of introducing a 2-form in a theory. Another way is to introduce a 2-form field interacting with Dirac spinors. This 2-form must involve the derivative of the 1-form gauge field i.e.\ $\ud{\cal A}$. Let us call this new field $\mho$. Then we can construct a Lagrangian density of the form
\begin{equation}\label{l7}
\begin{split}
{\cal L}_7 &=  - \frac{1}{4}\, {\cal F}_{\mu\nu}\cdot {\cal F}^{\mu\nu} + \bar\psi (i\,\gamma^\mu D_\mu - m)\psi + i\bar \psi\, \Sigma^{\mu\nu}\mho_{\mu\nu}\,\psi \\
&\quad {} + \frac{1}{2}\, {(B\cdot B + \bar B \cdot \bar B)} + B\cdot \partial_\mu {\cal A}^\mu -\,i\,\partial_\mu\bar C \cdot D^\mu C,
\end{split}
\end{equation}
where $\Sigma^{\mu\nu} = \frac{1}{2}\{\gamma^\mu, \,\gamma^\nu\}$ is the spin-connection. Exploiting the Curci-Ferrari condition, $B + \bar B + i\,(\bar C \times C) = 0$, a coupled but equivalent Lagrangian density can be written as
\begin{equation}\label{l8}
\begin{split}
{\cal L}_8 &=  - \frac{1}{4}\, {\cal F}_{\mu\nu}\cdot {\cal F}^{\mu\nu} + \bar\psi (i\,\gamma^\mu D_\mu - m)\psi + i\bar \psi\, \Sigma^{\mu\nu}\mho_{\mu\nu}\,\psi \\
&\quad {} + \frac{1}{2}\, {(B\cdot B + \bar B \cdot \bar B)} - \bar B\cdot \partial_\mu {\cal A}^\mu -\,i\,D_\mu\bar C \cdot \partial^\mu C.
\end{split}
\end{equation}
Except the term involving $\mho$, the other terms in the Lagrangian densities \eqref{l7} and \eqref{l8} are invariant under the usual $\mathrm{SU}(N)$ gauge symmetry transformations:
\begin{equation}\label{gt1}
\begin{aligned}
&\psi(x) \to \psi'(x) = U(x)\,\psi (x), \qquad \bar\psi (x) \to \bar\psi' (x) = \bar\psi(x)\, U^\dagger (x), \\
&{\cal A} \to {\cal A'} =  U{\cal A}U^\dagger + \phi,\qquad {\cal F}' = U{\cal F}U^\dagger,\qquad \phi = i\, \ud U\,U^\dagger.
\end{aligned}
\end{equation}
The 2-form $\mho$ here is interacting with the Dirac spinors and at this stage there is no reason to ignore its interaction with the 1-form gauge connection $\cal A$. As is obvious, for the interaction term, and hence the Lagrangian densities \eqref{l7} and \eqref{l8}, to be gauge invariant, $\mho\psi$ must transform as $(\mho\psi)' =  U\,(\mho\psi)$. The suitable combination of 2-form and 1-form connections, $\mho = {\cal B} - i\, \ud{\cal A}$, meets this requirement leading to the transformation of the $\mathcal{B}$-field found in literature \cite{Lahiri2002a, Lahiri2002b, Lahiri1997}, as we demonstrate now. We see that
\begin{equation} 
(\mho\psi)' = ({\cal B}' - i\,\ud{\cal A}') \psi' = [{\cal B}'U - i\,\ud(U{\cal A}U^\dagger + \phi)\,U] \psi,
\end{equation}
where we have made use of \eqref{gt1} in the second step. Equating this with
\begin{equation}
U\,(\mho\psi) = U\,({\cal B} - i\,\ud{\cal A})\,\psi = [U{\cal B} - i\,U\,\ud{\cal A}]\psi,
\end{equation}
we get ${\cal B}'U - i\,\ud(U{\cal A}U^\dagger + \phi)\,U = U{\cal B} - i\,U\,\ud{\cal A}$ which yields the gauge transformation of the $\mathcal{B}$-field:
\begin{equation}\label{gtb}
{\cal B}' = U{\cal B}U^\dagger + \phi\wedge U{\cal A}U^\dagger + U{\cal A}U^\dagger\wedge\phi + \phi\wedge\phi,
\end{equation}
where we have used $\ud\phi + i\,\phi\wedge \phi = 0$, which follows from the unitarity condition $UU^\dagger = U^\dagger U = 1$.    

Now we include the kinetic term for the 2-form fields and also the interaction between the 1-form and the 2-form in the Lagrangian density:
\begin{equation}\label{l9}
\begin{split}
{\cal L}_9 &= \frac{1}{12}\, {\cal W}_{\mu\nu\lambda}\cdot{\cal W} ^{\mu\nu\lambda} - \frac{1}{4}\, {\cal F}_{\mu\nu}\cdot {\cal F}^{\mu\nu} + \bar\psi (i\,\gamma^\mu D_\mu + i\,\Sigma^{\mu\nu}\mho_{\mu\nu} - m)\,\psi \\
&\quad {} + \frac{1}{2}\, {(B\cdot B + \bar B \cdot \bar B)} + B\cdot \partial_\mu {\cal A}^\mu -\,i\,\partial_\mu\bar C \cdot D^\mu C,
\end{split}
\end{equation}
where
\begin{equation}\label{w}
{\cal W} = \ud{\cal B} + i\,(\ud{\cal F} + \mathcal{A} \wedge {\cal B} - {\cal B}\wedge {\cal A})
\end{equation}
 is the usual field strength 3-form, which in component form reads
\begin{equation}
\begin{split}
{\cal W}_{\mu\nu\lambda} &= \partial_\mu{\cal B}_{\nu\lambda} + i\,\mathcal{A}_\mu \times \partial_{[\nu} \mathcal{A}_{\lambda]} - \mathcal{A}_\mu\times {\cal B}_{\nu\lambda} + \text{cyclic terms}\\
&=  \partial_\mu{\cal B}_{\nu\lambda} + i\,  ({\cal B}_{\mu\nu} - i\,\partial_{[\mu} \mathcal{A}_{\nu]}) \times \mathcal{A}_\lambda + \text{cyclic terms} \\
&= \partial_\mu {\cal B}_{\nu\lambda} + i\, \mho_{\mu\nu}\times \mathcal{A}_\lambda + \text{cyclic terms}.
\end{split}
\end{equation}
A coupled but equivalent Lagrangian density can be written, using the Curci-Ferrari condition, as
\begin{equation}\label{l10}
\begin{split}
{\cal L}_{10} &= \frac{1}{12}\, {\cal W}_{\mu\nu\lambda}\cdot{\cal W} ^{\mu\nu\lambda} - \frac{1}{4}\, {\cal F}_{\mu\nu}\cdot {\cal F}^{\mu\nu} + \bar\psi (i\,\gamma^\mu D_\mu + i\,\Sigma^{\mu\nu}\mho_{\mu\nu} - m)\,\psi \\
&\quad {} + \frac{1}{2}\, {(B\cdot B + \bar B \cdot \bar B)} - \bar B\cdot \partial_\mu {\cal A}^\mu -\,i\,D_\mu\bar C \cdot \partial^\mu C.
\end{split}
\end{equation}
The transformation of the 3-form ${\cal W}$ follows from the definition \eqref{w} and Eqs.~\eqref{gt1} and \eqref{gtb}. We provide some intermediate steps here.
\begin{gather}
\label{w01}\begin{split}
\ud{\cal B}' &= \ud \left( U {\cal B} U^\dagger + \phi \wedge U{\cal A} U^\dagger + U {\cal A}U^\dagger \wedge \phi + \phi \wedge \phi  \right) \\
&=  \ud U \wedge {\cal B}\,U^\dagger + U \,\ud{\cal B} U^\dagger + U \,{\cal B}\wedge \ud U^\dagger - i\,\ud U \wedge \ud U^\dagger \wedge U{\cal A}\,U^\dagger \\
&\quad {} + i\, U\, \ud U^\dagger \wedge \ud U\wedge {\cal A}U^\dagger - i\,\ud U\wedge \ud{\cal A}\,U^\dagger - i\,U \ud{\cal A} \wedge \ud U^\dagger\\
&\quad {} - i\, U{\cal A}\wedge \ud U^\dagger \wedge \ud U \,U^\dagger + i\,U {\cal A}\,U^\dagger \wedge \ud U\wedge \ud U^\dagger,
\end{split} \\
\label{w02}\begin{split}
i\,\ud{\cal F}' &= i\,\ud (U{\cal F}U^\dagger) = i\,\ud \left( U (\ud{\cal A} + i\,{\cal A}\wedge{\cal A})U^\dagger\right) \\
&= i\,\ud U \wedge \ud{\cal A}\,U^\dagger + i\,U\, \ud{\cal A} \wedge \ud U^\dagger - \ud U\wedge {\cal A}\wedge {\cal A}\,U^\dagger \\
&\quad {} - U\,\ud{\cal A}\wedge {\cal A}\,U^\dagger + U\,{\cal A}\wedge \ud{\cal A}\,U^\dagger - U\,{\cal A} \wedge {\cal A}\wedge \ud U^\dagger,
\end{split} \\
\label{w03}\begin{split}
{\cal A}'\wedge{\cal B}' &= (U{\cal A}U^\dagger + \phi) \wedge (U{\cal B}U^\dagger + \phi\wedge U{\cal A}U^\dagger + U{\cal A}U^\dagger \wedge \phi + \phi \wedge \phi) \\
&= U\,{\cal A} \wedge{\cal B}\,U^\dagger + U{\cal A}U^\dagger \wedge \phi \wedge U{\cal A}\,U^\dagger + U\,{\cal A} \wedge{\cal A}\,U^\dagger \wedge \phi \\
&\quad {} + U{\cal A}U^\dagger \wedge \phi \wedge \phi + \phi \wedge U{\cal B}U^\dagger + \phi\wedge \phi\wedge U {\cal A}\,U^\dagger \\
&\quad {} + \phi\wedge U{\cal A}U^\dagger \wedge \phi + \phi\wedge\phi\wedge\phi,
\end{split}\\
\label{w04}\begin{split}
{\cal B}'\wedge {\cal A}' &= (U{\cal B}U^\dagger + \phi\wedge U{\cal A}U^\dagger + U{\cal A}U^\dagger \wedge \phi + \phi \wedge \phi) \wedge  (U{\cal A}U^\dagger + \phi) \\
&=  U\,{\cal B} \wedge {\cal A}\,U^\dagger + U{\cal B}U^\dagger \wedge \phi + \phi\wedge U{\cal A}\wedge {\cal A}\,U^\dagger \\
&\quad {} + \phi \wedge U {\cal A}U^\dagger \wedge \phi + U{\cal A}U^\dagger \wedge \phi \wedge U{\cal A}U^\dagger \\
&\quad {} + U{\cal A}U^\dagger \wedge \phi \wedge \phi + \phi\wedge \phi \wedge U{\cal A}U^\dagger + \phi \wedge \phi \wedge \phi.
\end{split}
\end{gather}
From \eqref{w03} and \eqref{w04} it follows that 
\begin{equation}
\begin{split}
i \left( {\cal A}'\wedge {\cal B}' - {\cal B}'\wedge {\cal A}'\right)
&= i\,U ({\cal A}\wedge {\cal A}  - {\cal B}\wedge {\cal A})\,U^\dagger + \ud U\wedge {\cal A}\wedge {\cal A}\,U^\dagger\\
&\quad {} - U {\cal B}\wedge \ud U^\dagger,
\end{split}
\end{equation}
which along with \eqref{w01} and \eqref{w02} finally gives the transformation of the 2-form $\mathcal{W}$:
\begin{equation}\label{gtw}
\begin{split}
{\cal W}' &= \ud{\cal B}' + i \left( \ud{\cal F}' + {\cal A}'\wedge {\cal B}' - {\cal B}'\wedge {\cal A}'\right) \\
&= U \left[ \ud{\cal B} + i \left( \ud{\cal F} + {\cal A}\wedge {\cal B} - {\cal B}\wedge {\cal A}\right)\right] U^\dagger \\
&= U\,{\cal W}\,U^\dagger.
\end{split}
\end{equation}
Thus the Lagrangian densities \eqref{l9} and \eqref{l10} respect the $\mathrm{SU}(N)$ gauge symmetries \eqref{gt1}, \eqref{gtb} and \eqref{gtw}.

In order to obtain the (anti-)BRST symmetry transformations for the $\cal B$-field, we again apply the superunitary operator approach:
\begin{equation}\label{b-upgrade}
\tilde{\mathcal{B}} = \tilde U{\cal B} \tilde U^\dagger + \tilde\phi \wedge \tilde U {\cal A} \tilde U^\dagger + \tilde U{\cal A}\tilde U^\dagger\wedge \tilde\phi+ \tilde\phi \wedge \tilde\phi, \quad \tilde\phi = i\, \tilde{\ud} \tilde{U}\, \tilde{U}^\dagger = -i\, \tilde{U}\, \tilde{\ud} \tilde{U}^\dagger.
\end{equation}
Being a 2-form, $\tilde{\mathcal{B}}$ can be written as
\begin{equation}\label{m}
\begin{split}
\tilde{\mathcal{B}} (x, \theta, \bar\theta) &= (\ud x^\mu \wedge \ud x^\nu) \mathcal{M}_{\mu\nu} (x, \theta, \bar\theta) + \cdots \\
&= \mathcal{M} (x, \theta, \bar\theta) + \cdots,
\end{split}
\end{equation}
where, on the right-hand side, we have written only the term which is relevant for our purpose.%
\footnote{Other terms hidden inside ``$\cdots$'' are those which involve $\ud x^\mu \wedge \ud \theta$, $\ud x^\mu \wedge \ud \bar\theta$, $\ud \theta \wedge \ud \theta$, $\ud \bar\theta \wedge \ud \bar\theta$ or $\ud \theta \wedge \ud \bar\theta$.}
It is easy to see that if we use \eqref{m} on the left-hand side of \eqref{b-upgrade}, then we get
\begin{equation}\label{m-upgrade}
\mathcal{M} = \tilde U{\cal B} \tilde U^\dagger + \phi' \wedge \tilde U {\cal A} \tilde U^\dagger + \tilde U{\cal A}\tilde U^\dagger\wedge \phi' + \phi' \wedge \phi',
\end{equation}
where $\phi' = i\, {\ud} \tilde{U}\, \tilde{U}^\dagger = -i\, \tilde{U}\, {\ud} \tilde{U}^\dagger$. With the help of \eqref{u-na}, we now compute the right-hand side of \eqref{m-upgrade}. We find
\begin{gather}
\tilde{U} \mathcal{B} \tilde{U}^\dagger = {\cal B} + i\,\theta \,[{\cal B}, \bar C] + i\,\bar\theta\, [{\cal B}, C] + \theta\bar\theta \,\left( [B, {\cal B}] + \lbrace [C, {\cal B}], \bar C \rbrace \right), \\
\phi' = \theta \ud \bar C + \bar\theta \ud C + i \theta\bar\theta (\ud \mathcal{B} + \bar C \ud C + \ud C \bar C),\\
\begin{split}
\phi' \wedge \tilde{U} \mathcal{A} \tilde{U}^\dagger &= \theta \,(\ud\bar C \wedge {\cal A}) + \bar\theta \,(\ud C \wedge{\cal A}) \\
&\quad {} + i\,\theta\bar\theta \,\left(\ud B \wedge{\cal A} + \lbrace \ud C \wedge {\cal A}, \bar C\rbrace + \ud\bar C \wedge [ C, {\cal A}] \right),
\end{split} \\
\begin{split}
\tilde{U} \mathcal{A} \tilde{U}^\dagger \wedge \phi' &= \theta\, ({\cal A}\wedge \ud\bar C) + \bar\theta \,({\cal A}\wedge \ud C) \\
&\quad {} + i\,\theta\bar\theta \,\left({\cal A}\wedge \ud B + \lbrace {\cal A} \wedge \ud C, \bar C\rbrace + [{\cal A}, C] \wedge \ud\bar C \right), 
\end{split}\\
\begin{split}
\phi' \wedge \phi' = \theta\bar\theta \,(\ud C \wedge \ud\bar C - \ud\bar C\wedge \ud C).
\end{split}
\end{gather}
We thus find the final relation between $\mathcal{M}$ and $\cal B$:
\begin{equation}\label{b-map}
\begin{split}
\mathcal{M} &= {\cal B} + \theta \left( \ud\bar C\wedge {\cal A} + {\cal A}\wedge \ud\bar C + i\,[{\cal B}, \bar C] \right)
 + \bar\theta \left( \ud C\wedge {\cal A} + {\cal A}\wedge \ud C + i\,[{\cal B}, C ] \right) \\
&\quad {} + \theta\bar\theta \big( [B, {\cal B}] + \lbrace \left([C, {\cal B}] + i\,\ud C\wedge {\cal A} + i\, {\cal A}\wedge \ud C\right), \bar C\rbrace \\
&\qquad\qquad {} + i\, \mathcal{A}\wedge \ud B + i\, \ud B \wedge \mathcal{A} + DC \wedge \ud\bar C - \ud\bar C\wedge DC \big).
\end{split}
\end{equation}
Comparison of this equation with the standard expansion
\begin{equation}
\mathcal{M} = {\cal B} + \theta\,(s_{ab}{\cal B}) + \bar\theta\,(s_b {\cal B}) 
+ \theta\bar\theta \,(s_b\,s_{ab}{\cal B})
\end{equation}
yields the following (anti-)BRST transformations of the $\mathcal{B}$-field:
\begin{gather}
\label{sbb}
s_b\,{\cal B} = \ud C\wedge {\cal A} + {\cal A}\wedge \ud C + i\,[{\cal B}, C ], \\
\label{sabb}
s_{ab}\,{\cal B} = \ud\bar C\wedge {\cal A} + {\cal A}\wedge \ud\bar C + i\,[{\cal B}, \bar C], \\
\label{sbabb}
\begin{split}
s_b s_{ab}\,{\cal B} &= [B, {\cal B}] + \lbrace \left([C, {\cal B}] + i\,\ud C\wedge {\cal A} + i\, {\cal A}\wedge \ud C\right), \bar C\rbrace \\
&\quad {} + i\, \mathcal{A}\wedge \ud B + i\, \ud B \wedge \mathcal{A} + DC \wedge \ud\bar C - \ud\bar C\wedge DC.
\end{split}
\end{gather}
As a consistency check, we now use \eqref{sbb}, \eqref{sabb} and \eqref{brst-trans-na} to compute $s_b\,s_{ab}{\cal B}$. Equation \eqref{sbabb} is then reproduced.

In analogy with the 1-form gauge theories, we can now introduce a new 2-form covariant derivative $\mho_\mathcal{B}$. For any vector in Lie space, $P = P^a T^a$, this is given as
\begin{equation}
\mho_\mathcal{B} P = i\,[{\cal B}, P ] + \ud P\wedge {\cal A} + {\cal A}\wedge \ud P,
\end{equation}
while for spinor $\psi$, which is  a scalar in Lie space, this would be
\begin{equation}
\mho_\mathcal{B} \psi = i \mathcal{B} \psi + \mathcal{A} \wedge \ud \psi.
\end{equation}
Then \eqref{sbb}, \eqref{sabb} and \eqref{sbabb} can be rewritten as
\begin{gather}
s_b\,{\cal B} = \mho_\mathcal{B} C, \\
s_{ab}\,{\cal B} = \mho_\mathcal{B} \bar C, \\
s_b s_{ab}\,{\cal B} = i \mho_\mathcal{B} B + i \{\bar C, \mho_\mathcal{B} C\} + D C \wedge \ud \bar C - \ud \bar C \wedge D C.
\end{gather}

Before we conclude, we propose a conjecture for the gauge transformation of any $p$-form in a theory involving the 1-form and $p$-form gauge fields:
\begin{equation}
\begin{split}
K' &= UKU^\dagger + \chi\wedge \phi\wedge\phi\wedge\phi\wedge \ldots \wedge\phi + \phi\wedge \chi\wedge\phi\wedge\phi\wedge\ldots \wedge\phi + \ldots \\ 
&\quad {} + \phi\wedge \phi\wedge\phi\wedge \ldots \wedge\phi\wedge\chi + \phi\wedge\phi\wedge\phi\wedge\ldots \wedge\phi, 
\end{split}
\end{equation}
where $\chi = U{\cal A}U^\dagger$ and $\phi = i\,\ud U\,U^\dagger$. For 1-form this gives
\begin{equation}
{\cal A'} = \chi + \phi,
\end{equation}
while for the 2-form,
\begin{equation}
{\cal B'} = U{\cal B}U^\dagger + \chi\wedge \phi + \phi \wedge \chi + \phi\wedge\phi.
\end{equation}
Similarly, the transformation of 3-form follows:
\begin{equation}
{\cal W'} = U{\cal W}U^\dagger + \,\chi\wedge \phi\wedge\phi  + \,\phi\wedge \chi\wedge\phi 
+ \,\phi\wedge \phi\wedge\chi +\, \phi\wedge\phi\wedge\phi.
\end{equation}


\section{\label{sec:conclu}Results and discussion}

In this article, we started with a review of the superunitary operator formalism to obtain the (anti-)BRST transformations for 1-form gauge theories. We discussed the geometrical origin of the gauge fields as an interaction of the charged particle with the background spacetime. This approach is very useful as we can see how automatically the electromagnetic field comes in picture as a geometric phenomenon. This idea can be generalised to get the higher-form gauge fields too.

Exploiting the celebrated horizontality condition and the gauge-invariant restrictions, we computed explicitly the superunitary operator and obtained the (anti-)BRST transformations for the gauge field, the matter field and the (anti-)ghost field. We explored both the Abelian as well as the non-Abelian cases in our endeavour. As expected, the superunitary operator for the non-Abelian case, could be reduced to that in the Abelian case by using the Curci-Ferrari condition. 

We introduced a new Lagrangian with a coupling of matter fields not only with 1-from background field but also with a 2-form field. To restore the gauge symmetry in the theory it was necessary to introduce a new covariant derivative involving a 2-form gauge field $\mathcal{B}$. The two gauge fields further couple mutually. The transformation of $\mathcal{B}$ followed naturally from the definition of the 2-form covariant derivative. The same unitary operator, which yielded the transformations for the 1-form gauge and matter fields, was used to obtain the transformations for the 2-form gauge field as well. 

Generalising the idea of the unitary operator for any object---point or extended---in arbitrary background, we proposed a conjecture for the transformation of the associated $p$-form gauge field interacting with the Dirac spinors. This idea is very important as it equally works well for any supersymmetric gauge theory.


\section*{Acknowledgements}

Part of this work was done at the Department of Physics and the Centre of Theoretical Physics, Jamia Millia Islamia, New Delhi. D.S. gratefully acknowledges help from stimulating discussions with Pankaj Sharan. He is also thankful to K.N. Pathak and M.M. Gupta at Department of Physics, Panjab University, for providing facilities which helped to complete this work.



\end{document}